\documentclass[twocolumn, apj]{emulateapj}
\usepackage{amsmath}
\newcommand{\me}{\mathrm{e}} 
\newcommand{\dif}{\mathrm{d}}
\newcommand{\degree}{\ensuremath{^\circ}}

\shorttitle{3D Magnetothermal Instability}
\shortauthors{Parrish \& Stone}

\begin{document}

\title{Saturation of the Magnetothermal Instability in Three Dimensions}

\author{Ian J. Parrish and James M. Stone\altaffilmark{1}}
\affil{Department of Astrophysical Sciences, Princeton University, Princeton, NJ 08544}

\altaffiltext{1}{Program in Applied and Computational Mathematics, Princeton University, Princeton, NJ, 08544}

\begin{abstract}
In dilute astrophysical plasmas, thermal conduction is primarily  along magnetic field lines, and therefore highly anisotropic.  As a result, the usual convective stability criterion is modified from a condition on entropy to a condition on temperature.  For small magnetic fields or small wavenumbers, instability occurs in any atmosphere where the temperature and pressure gradients point in the same direction.  We refer to the resulting convective instability as the magnetothermal instability (MTI).  We present fully three-dimensional simulations of the MTI and show that saturation results in an atmosphere with different vertical structure, dependent upon the boundary conditions. When the temperature at the boundary of the unstable layer is allowed to vary, the temperature gradient relaxes until the unstable region is almost isothermal.   When the temperature at the boundary of the unstable region is fixed, the magnetic field is reoriented to an almost vertical geometry as a result of buoyant motions. This case exhibits more vigorous turbulence. In both cases the resulting saturated heat flux is almost one-half of the value expected if the conduction were purely isotropic, $\widetilde{Q}\sim\chi\Delta T/L $, where $\chi$ is the thermal conductivity, $\Delta T$ is the fixed temperature drop across the simulation domain, and $L$ is the temperature gradient scale length.  The action of the MTI results in dynamical processes that lead to significant transport perpendicular to the initial direction of the magnetic field.  The resulting magnetoconvection in both cases amplifies the magnetic field until it is almost in equipartition with sustained subsonic turbulence.  These results are relevant to understanding measurements of the temperature profiles of the intracluster medium of clusters of galaxies as well as the structure of radiatively inefficient accretion flows.
\end{abstract}

\section{Introduction} \label{sec:intro}
In many astrophysical plasmas the collision frequency is much smaller than the Larmor frequency, so that particles spiral around the magnetic field lines for very long distances between collisions.  As a result, transport of heat and momentum by thermal conduction and viscosity is highly anisotropic with respect to the magnetic field orientation.  In a dilute magnetized plasma, electron thermal conductivity parallel to the magnetic field is many orders of magnitude larger than both the perpendicular electron conductivity and either component of the ion conductivity provided $\omega \le \nu_{ee} \ll \Omega_e$, where $\nu_{ee}$ is the electron collision frequency, $\Omega_e$ is the electron gyrofrequency, and $\omega$ is the frequency of a physical process of interest.   In this regime the plasma is decribed by the equations of magnetohydrodynamics (MHD) with the addition of Braginskii anisotropic transport terms \citep{brag65}.  A Braginskii description is valid for describing processes where the mean free path is a substantial fraction of the size of the system, but does not permit an analysis of processes that occur on lengthscales shorter than the mean free path, e.g. Landau damping.  

These properties can have profound effects on the resulting plasma dynamics.  A striking example is the magnetothermal instability (MTI) \citep{bal00, ps05}.
The magnetothermal instability occurs as a result of heat flow parallel to the magnetic field in a statified atmosphere.  The criterion for convective stability is modified from the well-known Schwarzschild criterion $\left(\partial S/\partial z > 0\right)$, where the entropy, $S=p\rho^{-\gamma}$, to the Balbus criterion $(\partial \ln T/\partial z > 0)$, where $p$ and $T$ are the pressure and temperature respectively.  Atmospheres that have temperature (as opposed to entropy) profiles decreasing upwards are unstable to the MTI. In this situation the MTI is able to use the temperature gradient as a source of free energy for the instability.  The growth rate and saturation of this instability in two dimensions has been explored in \citet{ps05}, hereafter PS.  In PS we demonstrated that computationally measured growth rates match theoretical estimates from a WKB theory for a variety of boundary and initial conditions.  In addition, we showed that MTI-unstable plasmas can produce vigorous convective motions and a significant advective heat flux.  The saturated state for an adiabatic atmosphere in 2D was shown to result in an isothermal atmosphere.

We are motivated to study the nonlinear regime of the magnetothermal instability in three dimensions for several reasons.  First, the saturation mechanism in three dimensions may be different from two dimensions.  Second, there is a potential for a magnetic dynamo in three dimensions.  Finally, the nature of convection is known to be intrinsically different in three dimensions.  

In this paper we present the saturation and heat transport properties of the MTI in three dimensions.  We find the instability is capable of generating a magnetic dynamo that greatly amplifies the initial field until it is roughly in equipartition with the fluid turbulent kinetic energy.  To estimate the efficiency of heat transport, we can compare to a fiducial heat flux that would be expected if the conduction were purely isotropic,  $\widetilde{Q}\sim\chi\Delta T/L $, where $\chi$ is the thermal conductivity, $\Delta T$ is the fixed temperature drop across the simulation domain, and $L$ is the temperature gradient scale length.  We find that the saturated state of the atmosphere varies depending on the boundary condition, but is consistently able to transport a large fraction (30-50\%) of the fiducial heat flux.  The understanding of the three dimensional behavior of the MTI is likely to be very applicable to several astrophysical systems, such as the intracluster medium of clusters of galaxies, the atmospheres of neutron stars, and radiatively inefficient accretion flows.  

This paper is organized as follows.  In \S \ref{sec:method} we describe our numerical method and the physical mechanism responsible for the MTI.  In \S \ref{sec:saturation} we briefly comment on the linear properties of the MTI before focusing on the three-dimensional saturation mechanism in three different principal cases.  In \S \ref{sec:apps} we discuss several potential applications of the MTI.  Finally, in \S \ref{sec:conclusion} we summarize our results and discuss future work.  
\section{Method and Physics of the MTI} \label{sec:method}
\subsection{Equations of MHD with Anisotropic Heat Conduction} \label{subsec:method:MHDeqn}
We solve the usual equations of ideal MHD with the addition of a vector heat flux, $\textbf{\em Q}$, and a vertical gravitational acceleration $\textbf{\em g} = -g_0 \boldsymbol{\hat{z}}$, 
\begin{equation}
\frac{\partial \rho}{\partial t} + \boldsymbol{\nabla}\cdot\left(\rho \boldsymbol{ v}\right) = 0,
\label{eqn:MHD_continuity}
\end{equation}
\begin{equation}
\frac{\partial(\rho\boldsymbol{v})}{\partial t} + \boldsymbol{\nabla}\cdot\left[\rho\boldsymbol{vv}+\left(p+\frac{B^{2}}{8\pi}\right)\mathbf{I} -\frac{\boldsymbol{BB}}{4\pi}\right] + \rho\boldsymbol{g}=0,
\label{eqn:MHD_momentum}
\end{equation}
\begin{equation}
\frac{\partial E}{\partial t} + \boldsymbol{\nabla}\cdot\left[\boldsymbol{v}\left(E+p+\frac{B^{2}}{8\pi}\right) - \frac{\boldsymbol{B}\left(\boldsymbol{B}\cdot\boldsymbol{v}\right)}{4\pi}\right] 
+\boldsymbol{\nabla}\cdot\boldsymbol{Q} +\rho\boldsymbol{g}\cdot\boldsymbol{v}=0,
\label{eqn:MHD_energy}
\end{equation}
\begin{equation}
\frac{\partial\boldsymbol{B}}{\partial t} + \boldsymbol{\nabla}\times\left(\boldsymbol{v}\times\boldsymbol{B}\right)=0,
\label{eqn:MHD_induction}
\end{equation}
where the symbols have their usual meaning. The total energy $E$ is given as
\begin{equation}
E=\epsilon+\rho\frac{\boldsymbol{v}\cdot\boldsymbol{v}}{2} + \frac{\boldsymbol{B}\cdot\boldsymbol{B}}{8\pi},
\label{eqn:MHD_Edef}
\end{equation}
and the internal energy, $\epsilon=p/(\gamma-1)$.  For this paper, we assume
$\gamma=5/3$ throughout.  Note that equations (\ref{eqn:MHD_continuity})--(\ref{eqn:MHD_energy}) are in conservative form with the exception of the gravitational source terms.  The induction equation (Eqn. [\ref{eqn:MHD_induction}]) requires special treatment in MHD to preserve the $\boldsymbol{\nabla}\cdot \boldsymbol{B} = 0$ constraint exactly.  

The heat flux, $\boldsymbol{Q}  = \boldsymbol{Q_{C}} + \boldsymbol{Q_{R}}$, 
where $\boldsymbol{Q_{C}}$, the anisotropic term is due to Coulombic conduction parallel to magnetic field lines.  Cross-field collisions could contribute to the isotropic term, but will be neglected in this paper due to their very small magnitude in the astrophysical situations of interest--the result of the gyroradius being much smaller than the mean free path.  
The isotropic heat flux, $\boldsymbol{Q_{R}}$ is due to radiative transport.  Then
\begin{equation}
\boldsymbol{Q}_{C} = - \chi_{C} \boldsymbol{\hat{b}\hat{b}}\cdot\boldsymbol{\nabla}T,
\label{eqn:coulombic}
\end{equation}
\begin{equation}
\boldsymbol{Q}_{R} = - \chi_{R} \boldsymbol{\nabla}T,
\label{eqn:radiative}
\end{equation}
where $\chi_{C}$ is the Spitzer Coulombic conductivity \citep{spitz62},
$\boldsymbol{\hat{b}}$ is a unit vector in the direction of the magnetic field, and
$\chi_{R} $ is the coefficient of isotropic conductivity.  For the purposes of the general study of the MTI, we consider both coefficients as free parameters.
\subsection{Physics of the MTI}
The physics of the MTI can be understood simply by considering a small fluid parcel in a dilute, equilibrium, convectively stable atmosphere where the temperature decreases with height \citep{ps06}. 
Imagine a parcel of fluid at height $z_1$ in an equilibrium state with the initial magnetic field perpendicular to the temperature gradient, say in the $\hat{x}$ direction.  
In this initial configuration $\boldsymbol{\hat{b}}\cdot\boldsymbol{\nabla} T = 0$, and no heat flows.  

Now, perturb the parcel of fluid to a height $z_2 = z_1 + \delta z$.  The displaced parcel adiabatically cools, but the fluid element is still connected by the magnetic field to its hotter, original depth.  The resultant upward heat transport along the field causes the displaced parcel to become buoyant.  To understand the instability criterion, we note that this discussion requires mechanical equilibrium (mediated by sound waves) to operate faster than thermal conduction \citep{cd06}.  The temperature of the displaced fluid parcel is that of the lower fluid element, but its pressure $P^* = P_2$.  In a similar manner to \citet{cd06}, if we equate these pressures, assuming $P=\rho T$, and utilize a Taylor series for $T_2$ about $z_1$, we find
\begin{equation}
\left(\rho^* - \rho_2\right) T_1 = \delta z \rho_2 \frac{\dif T}{\dif z},
\label{eqn:MTIphys}
\end{equation}
where $\rho^*$ is the density of the displaced element.  
If the density contrast, $(\rho^* - \rho_2) < 0$, then the fluid parcel is buoyant.  This realization implies that the instability criterion in its simplest form is simply
\begin{equation}
\frac{\dif T}{\dif z} < 0 \qquad \textrm{(instability)}.
\label{eqn:instab-simple}
\end{equation}

We can be more precise about the instability criterion and growth rates by following a WKB analysis performed in detail in \S 4 of \citet{bal00} and \S\S 3-4 of \citet{cd06}. We introduce the Brunt-V\"ais\"al\"a frequency, $N$,
\begin{equation}
N^{2} = -\frac{1}{\gamma \rho}\frac{\partial P}{\partial z}\frac{\partial \ln S}{\partial z},
\label{eqn:brunt-vaisala}
\end{equation}
the frequency of vertical oscillations in a stably
stratified atmosphere.  Preserving Balbus' notation, we also define two useful
quantities,
\begin{equation}
\chi_{c}' = \frac{\gamma - 1}{P}\chi_{c} \qquad \textrm{and} \qquad \chi' = \frac{\gamma - 1}{P}\left(\chi_{c}+\chi_{R}\right).
\label{eqn:cond_normalization}
\end{equation}
Using the Fourier convention for perturbations, $\exp{(\sigma t + ikx)}$,
and in the limit of a weak magnetic field,
the dispersion relation simplifies to the non-dimensionalized form
\begin{equation}
\left(\frac{\sigma}{N}\right)^{3} + \frac{1}{\gamma}\left(\frac{\sigma}{N}\right)^{2}\left(\frac{\chi' T k^2}{N}\right) +\left(\frac{\sigma}{N}\right) + \frac{\dif \ln T}{\dif \ln S}\left(\frac{\chi'_{c} T k^2}{N}\right) = 0.
\label{eqn:dispersionrelation}
\end{equation}

By analysis of the Routh-Hurwitz criterion, the general instability criterion is found to be
\begin{equation}
k^2v_{A}^{2} - \frac{\chi'_{C}}{\rho \chi'}\frac{\partial P}{\partial z}\frac{\partial \ln T}{\partial z} < 0,
\label{eqn:MTI_instability_criterion1}
\end{equation}
where the second term is the isothermal limit ($\gamma \rightarrow 1$) of the Brunt-V\"ais\"al\"a frequency.
Examination reveals that long wavelengths are the most easily destabilized and that strong magnetic fields can suppress the instability.  A large isotropic conductivity also can short-circuit the parallel heat conduction and suppress the instability.  In the limit of infinite wavelength or very weak magnetic field, the instability criterion reduces to the heuristically derived result in equation (\ref{eqn:instab-simple}).  In this limit, any atmosphere where temperature decreases with height is convectively unstable to the MTI.
\subsection{Initial Conditions} \label{subsec:Method:IC}
For testing linear growth rates, we use the same initial vertical equilibrium state  outlined in PS \S\S 2.2-2.4.  For studying the nonlinear saturation properties of the instability, we construct an atmosphere similar to \S 5.2 of PS where an MTI-unstable layer is surrounded on top and bottom by convectively stable layers.  This construction prevents the formation of unresolved thermal boundary layers at the edge of the computational domain.  In particular, we begin with the \textit{Ansatz} that the atmosphere exists in a constant gravitational field, $\boldsymbol{g}(z) = -g_0 \boldsymbol{\hat{z}}$, in a plane parallel atmosphere.  We divide the vertical size, $L_z$, into three equal regions.  The bottom and top layers, region 1 ($0\le z\le L_z/3$) and region 3 ($2L_z/3\le z\le L_z$) respectively, are isothermal atmospheres.  In such a region, the pressure and density vary exponentially with the height.  For a region with temperature $T_0$, the atmosphere has solutions of
\begin{equation}
\rho(z) = \rho_0 \me^{-g_0 z/T_0},
\label{eqn:isothermal-rho}
\end{equation}
\begin{equation}
P(z) = P_0 \me^{-g_0 z/T_0},
\label{eqn:isothermal-P}
\end{equation}
where $P_0 = T_0 \rho_0$.  Simultaneously, we set the conductivity in this region to be purely isotropic with $\chi_R = \chi_{\textrm{iso},0}$.  As a consequence of this choice, the second term of equation (\ref{eqn:MTI_instability_criterion1}) is zero and the MTI is stabilized.  For simplicity, we choose the parameters at the bottom of the isothermal region to have $T_b = P_b = \rho_b = 1$.  We designate the parameters at top of region 1 with the subscript 1, e.g. $T_1$.  

In the middle region, which we refer to as region 2, we construct an atmosphere where the temperature decreases linearly with height as in the previous 2D study.  Thus our initial \textit{Ansatz} is a power law atmosphere in the region 
$ L_z/3 \le z \le 2 L_z /3 $:
\begin{equation}
T(z) = T_{1}\left[1-\frac{(z-L_z/3)}{z_0}\right],
\label{eqn:T_profile}
\end{equation}
\begin{equation}
\rho(z) = \rho_{1}\left[1-\frac{(z-L_z/3)}{z_0}\right]^{2},
\label{eqn:rho_profile}
\end{equation}
\begin{equation}
p(z) = p_1\left[1-\frac{(z-L_z/3)}{z_0}\right]^{3},
\label{eqn:P_profile}
\end{equation}
where $z_0$ is a constant.  For stability to adiabatic convection, we require that $0 < z_{0}^{-1} < 2/5 g_{0}$.  Our choice of $z_0 =3$ satisfies this constraint.  Region 2 has conductivity that is exclusively anisotropic with $\chi_C = \chi_{\textrm{aniso},0}$.  Thus region 2 is unstable to the MTI.  We designate the hydrodynamic parameters at the top of this region with the subscript 2, e.g. $T_2$.

The top layer of the atmosphere, region 3, is constructed in an exactly analogous manner to Region 1.   It is an isothermal atmosphere with constant temperature $T_t = T_3$.  The thermal conductivity is again set to be purely isotropic.  

The magnetic field is initialized only in Region 2.  By initializing the magnetic field only in the MTI-unstable region, its propagation can be used as a tracer of phenomena like penetrative convection and magnetic pumping into the stable regions.  To initialize the field with a fully three dimensional structure, we choose in the region $ L_z/3 \le z \le 2 L_z /3 $, 
\begin{equation}
\boldsymbol{B}(z) = \left\{ \begin{array}{ll}
B_0 \cos\left[\nu \pi \frac{\left(z - L_z/3\right)}{L_z}\right] & 
\boldsymbol{\hat{x}}\\
-B_0 \sin\left[\nu \pi \frac{\left(z - L_z/3\right)}{L_z}\right]&
 \boldsymbol{\hat{y}},
\end{array} \right.
\label{eqn:B_profile}
\end{equation}
where $\nu = 3$ is a half rotation of the helix and represents a net flux condition and $\nu = 6$ is a full rotation of the helix and represents a no net flux condition.  The coefficient $B_0 = 10^{-5}$ is chosen so that magnetic tension effects are unimportant in the linear regime.
  
\subsection{Dimensionless Analysis} \label{subsec:Method:dimensionless}
There are four characteristic time-scales of the problem.  The first is the sound crossing time of the domain
\begin{equation}
\tau_{s} = L/c_{s}, \qquad c_s^2 = \frac{\gamma P}{\rho}.
\label{eqn:tau_s}
\end{equation}
For our standard parameters, the adiabatic soundspeed, $c_{s} \approx 1.28$, and the domain size is, $L=0.1$.  Thus, the sound crossing time is a short $\tau_s \approx 7.8\times 10^{-2}$.  Likewise, we can define an Alfv\'{e}n crossing time as 
\begin{equation}
\tau_A = L/v_A, \qquad v_A^2 = \frac{B_0^2}{4\pi \rho}.
\label{eqn:tau_A}
\end{equation}
For a typical initial magnetic field of $B_0=10^{-5}$, $\tau_A \approx 10^4$, many orders of magnitude longer than the sound-crossing time.  Thus, we see that Alfv\'{e}n waves are ordered out of the problem, as one would expect for a high-$\beta$ plasma.  
The third timescale is that of the atmospheric oscillation time. This time-scale is simply the inverse of the Brunt-V\"ais\"al\"a frequency,
\begin{equation}
\tau_N = N^{-1},
\label{eqn:tau_N}
\end{equation}
where $N$ is defined in equation (\ref{eqn:brunt-vaisala}).  For the gradients in region 2, $N \approx 0.26$, which gives $\tau_N \approx 3.9$.  Finally, the last timescale of interest is the heat diffusion time across the simulation domain.  It is obtained by non-dimensionalizing the thermal conduction equation, resulting in a time-scale
\begin{equation}
\tau_{\chi} = \frac{L^2 \Delta T}{\chi_{C} \bar{T}},
\label{eqn:tau_chi}
\end{equation}
where $\Delta T$ is the temperature difference across the unstable region, and $\bar{T}$ is the average temperature.  For the anisotropic conductivity of $\chi_C = 10^{-4}$ and the standard atmosphere, the heat conduction timescale, $\tau_{\chi} \approx 1.0$.

The asymptotic ordering of these terms is instructive:
\begin{equation}
\tau_s \ll \tau_{\chi} \lesssim \tau_N \ll \tau_A.
\label{eqn:ordering}
\end{equation}
First, the sound crossing time is rapid compared to the buoyancy time.  Second, we find that thermal conduction is able to smooth the temperature differences on a timescale comparable to the buoyant time.  Finally, as previously noted, the magnetic field is not dynamically important. 
\subsection{Numerical Methods}
We use the 3D version of the Athena MHD code \citep{gs05,gs06} to simulate the MTI far into the nonlinear regime.  Athena is a higher order, fully unsplit Godunov method that combines corner tranport upwind (CTU) methods with constrained transport (CT) to evolve the equations of MHD while manifestly maintaining the divergence-free constraint of the magnetic field.  We utilize the 6 step CTU+CT algorithm.  Isotropic and anisotropic heat fluxes are calculated as face-centered, area-averaged quantities as described in detail in the appendix of PS.  By finite-differencing the anisotropic diffusion terms we are able to keep the ratio of parallel conductivity to numerical perpendicular conductivity to better than $10^3$ for an arbitrary field orientation.  This level of truncation error is more than sufficient to simulate the MTI.  Sharma and Hammett (2006, personal communication) have found that for certain discontinous test problems this differencing can lead to negative temperatures; however, for our applications the temperature gradients are continuous and we observe \textit{post hoc} that no such problems arise.  The heat transport algorithm is operator split from the MHD and sub-cycled when the heat conduction timestep is shorter than the MHD Courant condition.  Both the MHD and heat conduction algorithms are parallelized using MPI.

All of the non-linear layered runs in this paper are performed on a three-dimensional grid initialized as described in \S \ref{subsec:Method:IC} with gravity in the $\boldsymbol{\hat{z}}$ direction.  The size of the domain is $L_x = L_y = 0.1$ and $L_z = 0.2$.  Our standard numerical resolution is $64^2\times 128$, such that our zones are cubic.  We have performed some simulations at twice this resolution in each direction.  Our box size is small compared to the atmospheric scale height, $H \sim c_{s}^{2}/g \approx 1.64$, where $c_{s}\approx 1.28$ is the adiabatic sound speed.  Since the MTI is a local instability, the evolution of modes with wavelengths much smaller than the scale height are independent of the exact form of $g(z)$.  

We use periodic boundary conditions for the MHD variables and the temperature in the $x$ and $y$ directions.  In the vertical direction, we use a modified reflecting boundary condition in which all MHD variables are reflected except the internal energy.  Due to the non-zero gravitational source term at the boundary, we extrapolate the pressure across the boundary to ensure the pressure gradient balances gravity, and the boundary cell face does not undergo spurious acceleration.  For the temperature in the $z$-direction we implement a Dirichlet boundary condition; that is, the temperature is held fixed at the boundary and heat is permitted to flow as necessary.  We refer to this boundary condition as ``conducting boundaries.''  

To seed multiple modes of the MTI, we add small Gaussian white noise perturbations in all three velocity components in the unstable region.  The standard perturbation amplitude is $v_0 = 10^{-4}$, which is highly subsonic.  The standard initial magnetic field strength is $B_0/(4\pi)^{1/2} = 10^{-5}$.  The Alfv\'{e}n velocity is then roughly $v_A \sim 10^{-5}$ (varying slightly with density), which is similar in magnitude to the perturbation velocity, although this fact does not affect the instability, as was shown in \S \ref{subsec:Method:dimensionless}.  
\section{Saturation of the MTI in 3D} \label{sec:saturation}
We have tested our 3D numerical algorithms by confirming the linear growth rates of the MTI.  As expected, we find that the measured growth rates are very close to the ones predicted from the WKB theory, and no deviation is found in the saturation of single-mode perturbations from the two-dimensional results.  We therefore quickly turn our attention to the non-linear regime in 3D.

Table \ref{tab:runs} lists the runs that will be discussed in this paper.  All of them have the computational set-up described in \S \ref{subsec:Method:IC} with constant conductivities.  Runs F1 -- F3 will be our fiducial cases of interest and will be discussed in the following sections.  These runs have a fixed anisotropic thermal conductivity, $\chi_{\textrm{aniso},0} = 1 \times 10^{-4}$, in the unstable layer with varying isotropic conductivities in the stable layers.  Runs N1 -- N2 examine the effect of a net magnetic flux on the fiducial runs, and run N3 is a high resolution version of run F3.  Runs N4 -- N6 have a higher magnitude of conductivity and are interesting for comparison to the canonical cases.  It should be noted that not all of these latter runs have reached full saturation, therefore we report the parameters of the fiducial runs at an early time for comparison purposes. We realized that the saturation mechanism had two very different forms with a continuum of intermediate cases.  The variation in saturation mechanism is directly related to the ratio of the conductivity in the MTI-stable outer layers ($\chi_{\textrm{iso},0}$) to the conductivity in the central MTI-unstable layer ($\chi_{\textrm{aniso},0}$).  This condition is essentially a ratio of the applied heat flux across the unstable layer to the parallel conductivity.

\begin{deluxetable*}{lccccc}
\tablecolumns{6}
\tablecaption{Properties of Nonlinear Runs  \label{tab:runs}}
\tablewidth{0pt}
\tablehead{
\colhead{Run} &
\colhead{$\chi_{\textrm{iso},0}$} &
\colhead{$\chi_{\textrm{aniso},0}$} &
\colhead{$B_0/\left(4\pi\right)^{1/2}$} &
\colhead{Flux\tablenotemark{a}} &
\colhead{Resolution}
}
\startdata
F1........... & $1 \times 10^{-5}$ & $1 \times 10^{-4}$ &
$1 \times 10^{-5}$ & 0 & $64 \times 64 \times 128$ \\
F2........... & $1 \times 10^{-4}$ & $1 \times 10^{-4}$ &
$1 \times 10^{-5}$ & 0 & $64 \times 64 \times 128$ \\
F3........... & $1 \times 10^{-3}$ & $1 \times 10^{-4}$ &
$1 \times 10^{-5}$ & 0 & $64 \times 64 \times 128$ \\
N1........... & $1 \times 10^{-5}$ & $1 \times 10^{-4}$ &
$1 \times 10^{-5}$ & + & $64 \times 64 \times 128$ \\
N2........... & $1 \times 10^{-3}$ & $1 \times 10^{-4}$ &
$1 \times 10^{-5}$ & + & $64 \times 64 \times 128$ \\
N3........... & $1 \times 10^{-3}$ & $1 \times 10^{-4}$ &
$1 \times 10^{-5}$ & 0 & $128 \times 128 \times 256$ \\
N4........... & $5 \times 10^{-5}$ & $5 \times 10^{-4}$ &
$1 \times 10^{-5}$ & 0 & $64 \times 64 \times 128$ \\
N5........... & $5 \times 10^{-5}$ & $5 \times 10^{-4}$ &
$1 \times 10^{-5}$ & + & $64 \times 64 \times 128$ \\
N6........... & $5 \times 10^{-4}$ & $5 \times 10^{-5}$ &
$1 \times 10^{-5}$ & + & $64 \times 64 \times 128$ \\
\enddata
\tablenotetext{a}{'+' indicates net magnetic flux threading the domain; '0' indicates zero net magnetic flux.}
\end{deluxetable*}

\begin{deluxetable*}{lclccccccc} 
\tablecolumns{10}
\tablecaption{Saturation Properties in the Nonlinear Regime \label{tab:satprop}}
\tablewidth{0pt}
\tablehead{
\colhead{Run} &
\colhead{Case} &
\colhead{$t/\tau_s$\tablenotemark{(a)}} &
\colhead{$\delta \langle B\rangle^2$} &
\colhead{$\delta \langle B^2\rangle$} &
\colhead{RMS Mach} &
\colhead{$\langle V_A\rangle$} &
\colhead{$\langle\theta_B\rangle$} &
\colhead{$\left(\frac{\dif T}{\dif z}\right)_{\textrm{mid}}^{*}$\tablenotemark{(b)}} &
\colhead{$\langle Q\rangle/\widetilde{Q}$}
}
\startdata
F1......... & A & $2,050$ & 20.6 & 33.4 & $8.9\times 10^{-5}$ & $4.6\times 10^{-5}$ & $17.5\degree$ & 15.8\% & 27\%\\
 & & $4,430^*$ & 22.7 & 36.2 & $5.7\times 10^{-5}$ & $4.9\times 10^{-5}$ & $18.9\degree$ & 10.8\% & 49\%\\
F2......... & Int & $2,050$ & 26.7 & 45.1 & $6.6\times 10^{-4}$ & $5.3\times 10^{-5}$ & $19.8\degree$ & 46.9\% & 27\%\\
 & & $10,700^*$ & 186 & 298 & $6.1\times 10^{-4}$ & $1.4\times 10^{-4}$ & $36.6\degree$ & 46.0\% & 27\%\\
F3......... & B & $2,050$ & 71.6 & 121 & $1.8\times 10^{-3}$ & $8.6\times 10^{-5}$ & $25.3\degree$ & 90.7\% & 45\%\\
 & & $8,470^*$ & 5290 & 7760 & $1.5\times 10^{-3}$ & $7.5\times 10^{-4}$ & $54.9\degree$ & 89.1\% & 50\%\\
N1......... & A & $2,050$ & 74.2 & 116 & $2.2\times 10^{-4}$ & $8.8\times 10^{-5}$ & $27.1\degree$ & 14\% & 30\%\\
N2......... & B & $2,050$ & 379 & 615 & $1.9 \times 10^{-3}$ & $2.0 \times 10^{-4}$ & $38.7\degree$ & 90.7\% & 45\%\\
N3......... & B & $2,050$ & 1680 & 3480 & $6.1 \times 10^{-4}$ & $4.1\times 10^{-4}$ & $35.7\degree$ & 90.9\% & 42\%\\
N4......... & A & $2,050$ & 22.1 & 36.8 & $3.5\times 10^{-4}$ & $4.8\times 10^{-5}$ & $17.9\degree$ & 10.7\% & 44\%\\
N5......... & A & $2,050$ & 97.8 & 163 & $3.5\times 10^{-4}$ & $1.01\times 10^{-4}$ & $29.5\degree$ & 10.6\% & 45\%\\
N6......... & B & $2,050$ & 144 & 243 & $1.2\times 10^{-3}$ & $1.2\times 10^{-4}$ & $32.2\degree$ & 89.5\% & 41\%\\
\enddata
\tablenotetext{(a)}{The symbol $^*$ indicates that the run is considered to be approximately saturated.}
\tablenotetext{(b)}{The temperature gradient is given as a percentage of the initial temperature gradient in the central region.}
\end{deluxetable*}

\subsection{Case A: Temperature Gradient Relaxation}  \label{subsec:saturation:F1}
\begin{figure} 
\epsscale{1.1}
\plotone{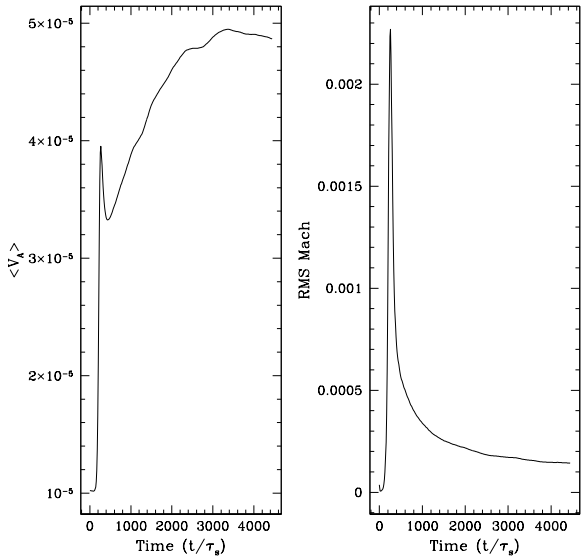}
\caption{Evolution of the Alfv\'{e}n velocity and RMS Mach number of run F1 averaged over the central MTI-unstable region.  The two are almost in equipartition.}\label{fig:F1-midavg}
\end{figure}

We begin our discussion of saturation by examing run F1, in which the conductivity in the isotropic, stable layers is an order of magnitude smaller than the anisotropic conductivity. 
\begin{figure*} 
\epsscale{0.7}
\plotone{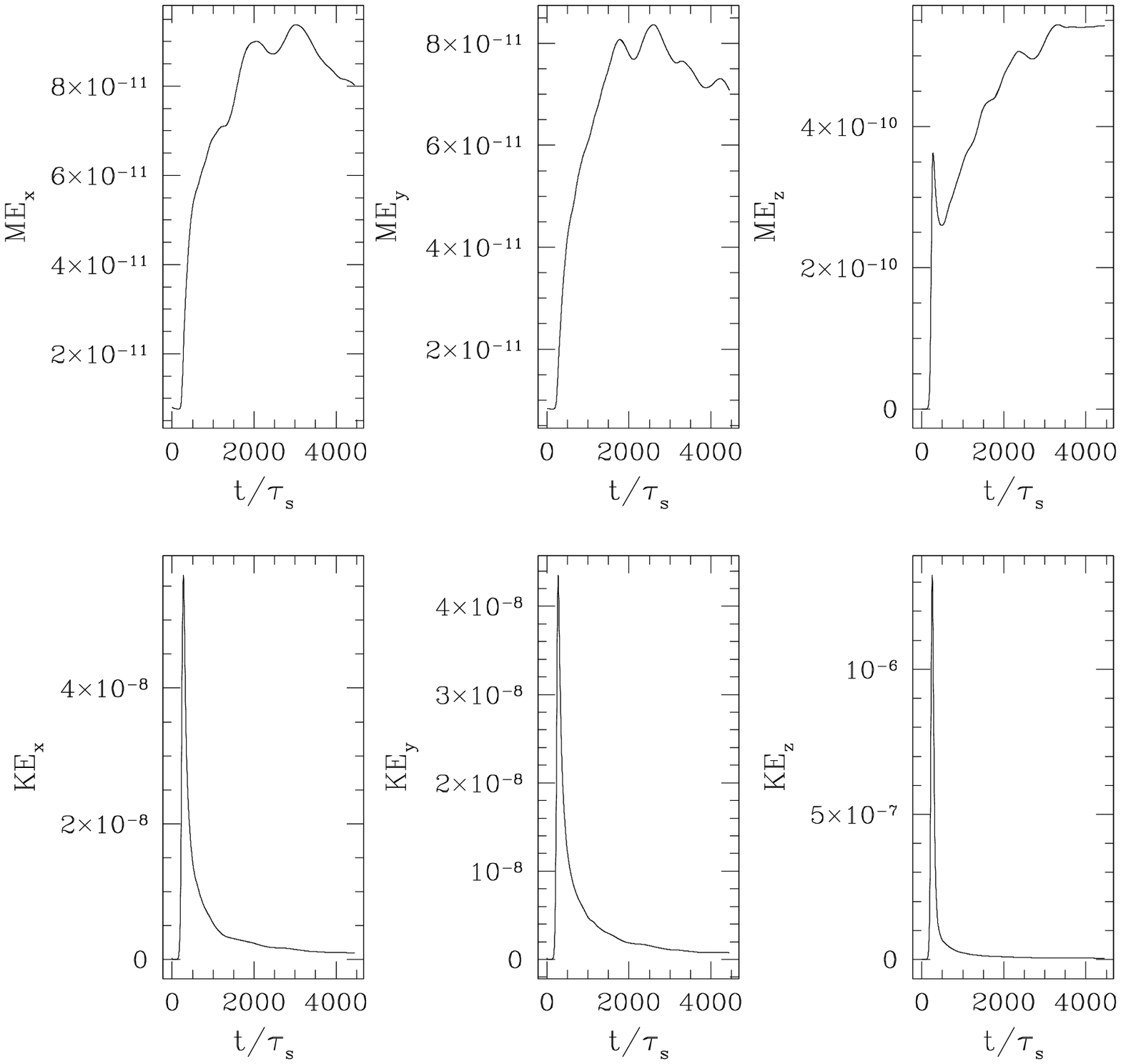}
\caption{Time evolution of the components of the volume-averaged magnetic and kinetic energy density for the entire domain for Run F1.}\label{fig:F1-time}
\end{figure*}
Figure \ref{fig:F1-time} shows the evolution of the volume-averaged components of the magnetic energy density and kinetic energy density over the entire domain, including the stable regions.  There is a very rapid spike in all the component of both magnetic and kinetic energy as the linear phase of the instability progresses.  The kinetic energy then decays to a relatively constant value as the magnetic field is amplified further.  Figure \ref{fig:F1-midavg} shows the increase of the volume--averaged Alfv\'{e}n velocity, $\langle V_A\rangle = \left\langle\left(B^2/\rho\right)^{1/2}\right\rangle$, as the magnetic field increases to $\langle V_A\rangle = 4.9\times 10^{-5}$. 
Meanwhile the RMS Mach number, $\left\langle v^2/c_s^2\right\rangle^{1/2}$, has saturated at a relatively constant $5.7\times 10^{-5}$ which corresponds to a vertical velocity, $\langle v_z\rangle = \langle c_s\rangle\langle \textrm{Ma} \rangle \approx 7.3\times 10^{-5}$.  
This Mach number represents highly subsonic turbulence. Thus, the magnetic energy density has been amplified such that it is almost in equipartition with the kinetic energy density.  We can quantitatively estimate the increase in magnetic energy density as listed in Table \ref{tab:satprop} with two similar but slightly different measures of the magnetic energy based on the square of the mean field or the mean of the squared field:
\begin{equation}
\delta\langle B \rangle^2 \equiv \frac{\langle B \rangle^2_{\textrm{fin}}}{\langle B \rangle^2_{\textrm{init}}} \qquad
\delta\langle B^2 \rangle \equiv \frac{\langle B^2 \rangle_{\textrm{fin}}}{\langle B^2 \rangle_{\textrm{init}}}.
\label{eqn:bamp}
\end{equation}
 At saturation, run F1 achieves a modest increase in these two measures of a factor of 20.6 and 33.4, respectively.  

\begin{figure} 
\epsscale{1.2}
\plotone{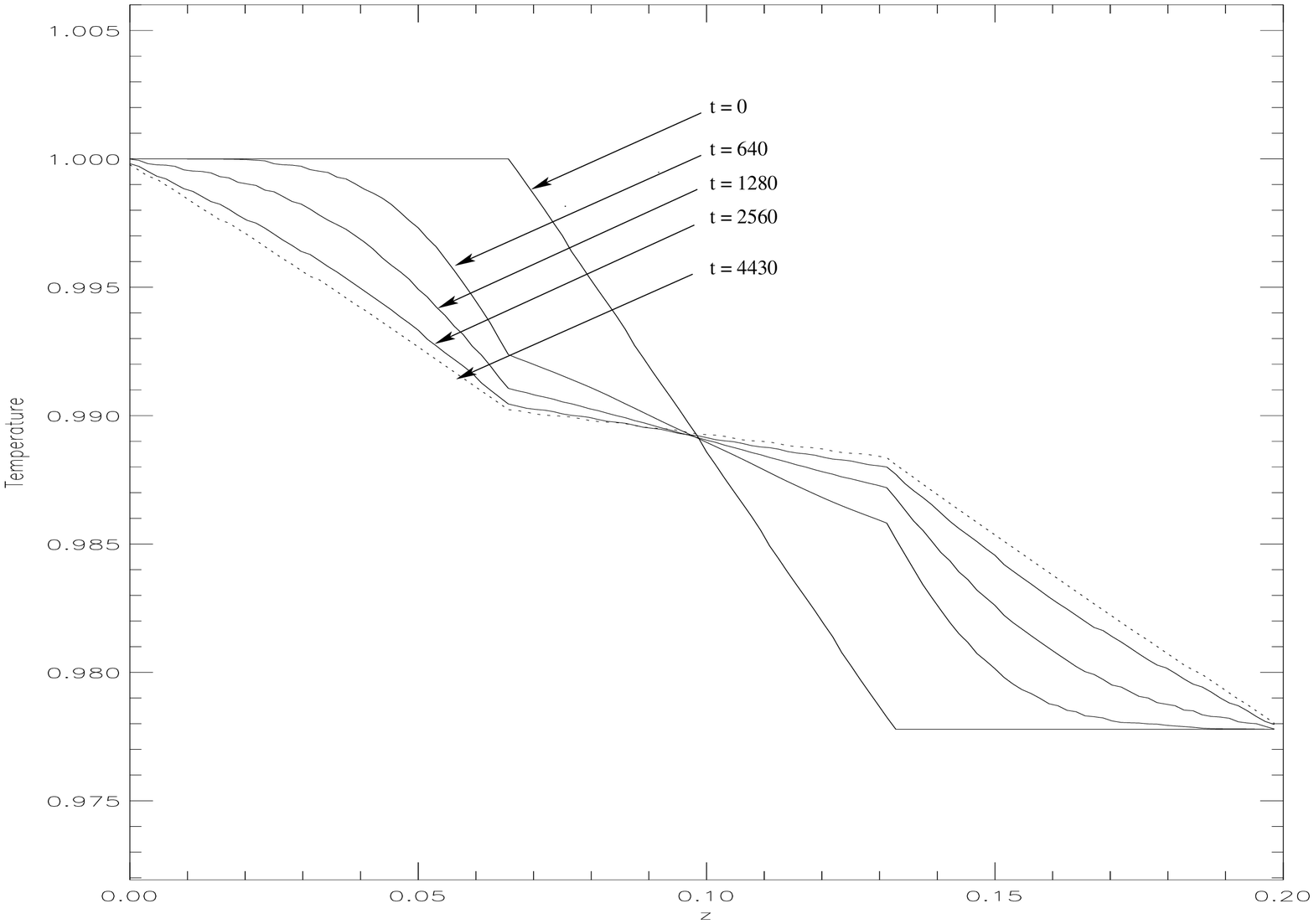}
\caption{Vertical profile of the horizontally-averaged temperature profile in run F1 at various times.  Times are labeled in terms of $\tau_s$.  The final, saturated state (dotted line) is almost isothermal in the central region.}\label{fig:F1-tavg}
\end{figure}

The evolution to saturation can be understood by examining the evolution of the horizontally-averaged temperature profile shown in Figure \ref{fig:F1-tavg}. 
The indicated times are shown non-dimensionalized to the sound crossing time. The temperature profile evolves very rapidly towards an essentially isothermal atmosphere in the MTI-unstable, central region.  The atmosphere in the stable layer slowly relaxes to an essentially linear temperature profile.  In this case, the temperature profile has relaxed to a quasi-isothermal state as seen previously in the 2D studies in PS with adiabatic boundary conditions.  Thus, the source of free energy, the temperature gradient, is quickly exhausted in the unstable region before it can be replenished by the slow heat conduction from the outer, stable layers.  We refer to this method of saturation as Case A: temperature gradient relaxation.  

\begin{figure} 
\epsscale{1.25}
\plotone{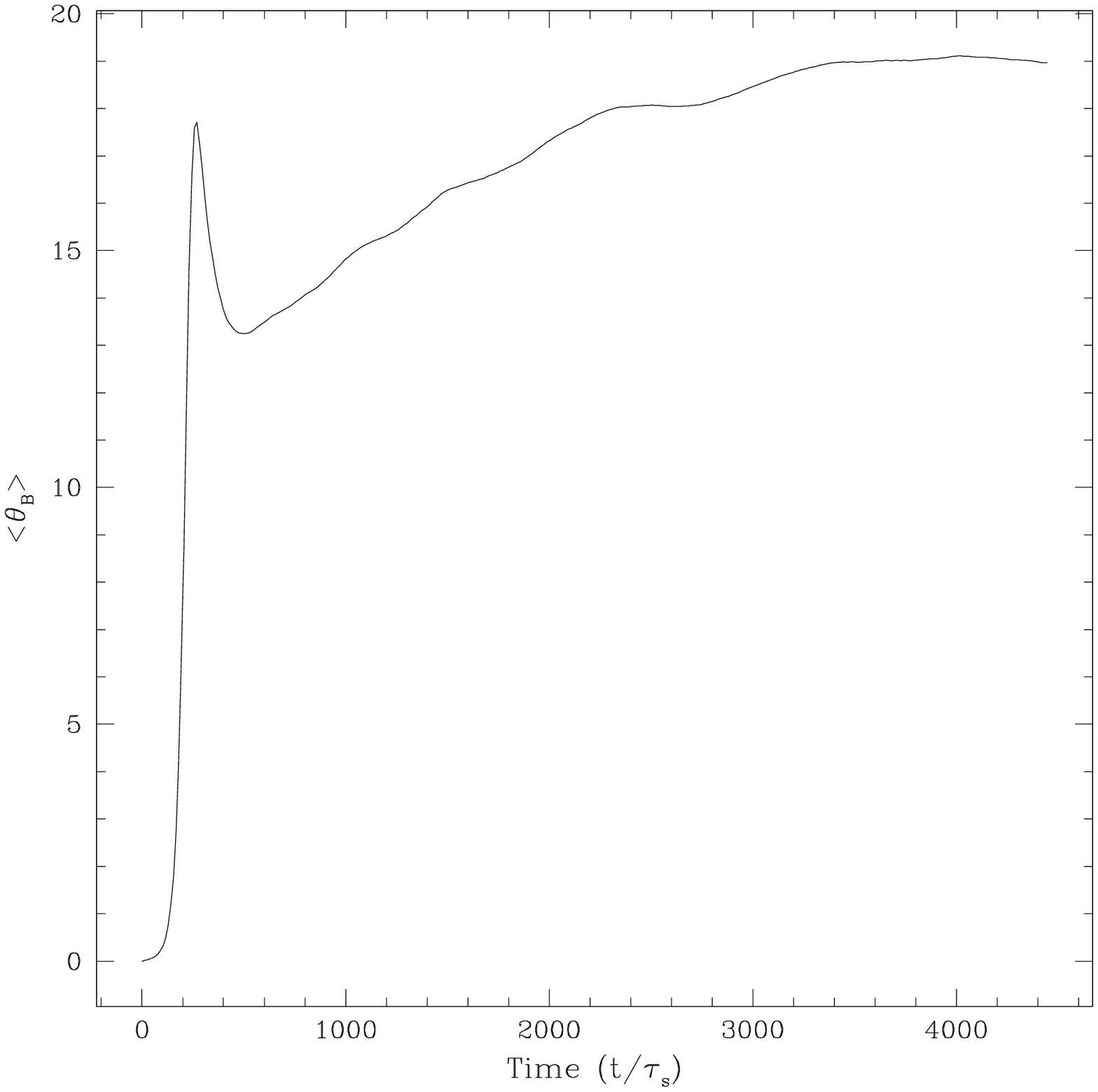}
\caption{Evolution of the average angle of the magnetic field with respect to the horizontal in degrees for run F1.  The sharp initial peak is followed by a much slower increase to its asymptotic value.}\label{fig:F1-btheta}  
\end{figure}

\begin{figure*} 
\epsscale{0.65}
\plotone{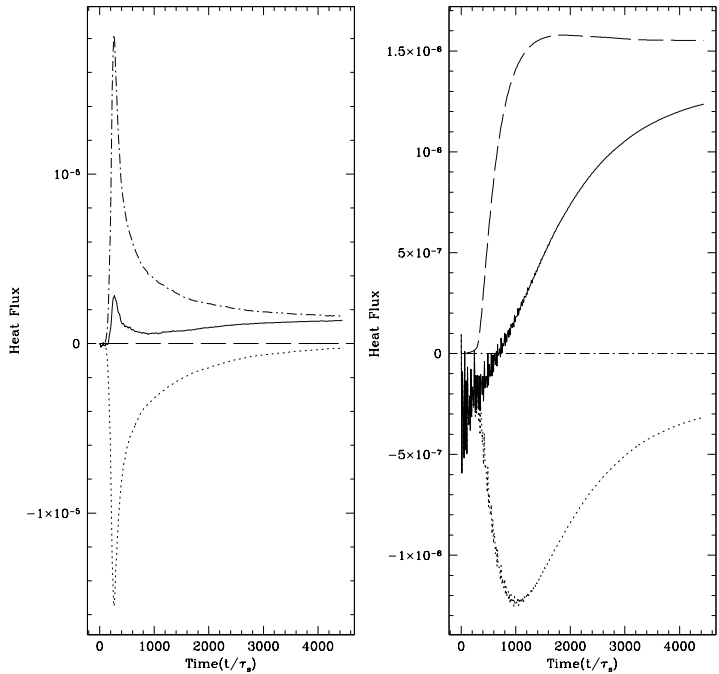}
\caption{Time evolution of the horizontally-averaged heat flux at the midplane (\textit{left panel}) and at 80\% height (\textit{right panel}) of the simulation domain in run F1.  The total heat flux (\textit{solid line}) is subdivided into conductive (\textit{dot-dash line}), radiative (\textit{long dashed line}), and advective (\textit{dotted line}) components.}\label{fig:F1-heat}  
\end{figure*}

An interesting insight into the saturated state can be gleaned from examining the average angle of the magnetic field with respect to the horizontal in the unstable layer, defined as
\begin{equation}
\langle\theta_B\rangle \equiv \left\langle \sin^{-1} \left(\frac{B_z}{|B|}\right)\right\rangle.
\label{eqn:thetab}
\end{equation}
This quantity is plotted in Figure \ref{fig:F1-btheta} for run F1.
The average angle of the magnetic field is initially zero---consistent with the initial field being perpendicular to the temperature gradient.  As the MTI grows exponentially, there is a rapid rise in the angle of the magnetic field at the same time as the rapid growth in the kinetic and magnetic energy densities.  After this initial rise in magnetic field angle, $\langle\theta_B\rangle$ slowly reaches an asymptotic value of $18.9\degree$.  The saturated state for this case is primarily reached by temperature gradient relaxation, although the effects of the rearrangement of the magnetic field are conflated with the former.  One can calculate a rough steady state heat flux in the anisotropic layer (region 2) as,
\begin{eqnarray}
Q_{\textrm{aniso}} &=& -\chi_{\textrm{aniso},0} \sin\langle\theta_B\rangle \left(\frac{\dif T}{\dif z}\right)_2 \nonumber \\
&=& -(1\times 10^{-4})(\sin 18.9\degree)(-0.0367) \nonumber \\
&\approx& \;\;\,\, 1.2\times 10^{-6},
\label{eqn:F1-anisoQ}
\end{eqnarray}
using the values from Table \ref{tab:satprop}.  
We can compare the measured Coulombic heat flux to a fiducial heat flux.  The heat flux that would occur if heat were to flow across the domain with purely isotropic conductivity for the given initial temperature gradient is defined as
\begin{equation}
\widetilde{Q} = - \min[\chi_{\textrm{aniso}}, \chi_{\textrm{iso}}]\left(\frac{\Delta T}{L}\right)_0,
\label{eqn:q-fiducial}
\end{equation}
which is for this run limited by the conductivity of the isotropic layer to be $\widetilde{Q} = 3.3\times 10^{-6}$.  Thus the steady-state anisotropic heat flux is a bit more than one-third of the fiducial value.  
The anisotropic heat flux can be compared to the steady state heat flux in the isotropic layers (regions 1 and 3) which is $Q_{\textrm{iso}} = \chi_{\textrm{iso},0}\left(\dif T/\dif z\right)_{1,3}\approx1.3\times10^{-6}$.  Thus, the steady state heat fluxes are roughly consistent.   In addition to the Coulombic (eq. [\ref{eqn:coulombic}]) and radiative (eq. [\ref{eqn:radiative}]) heat fluxes, we define an advective heat flux in the vertical direction, 
\begin{equation}
Q_{adv} \equiv V_z\epsilon,
\label{advective}
\end{equation}
where, $V_z$ is the vertical component of the fluid velocity, and $\epsilon=p/(\gamma
-1)$, is the internal energy.
The heat fluxes for this run are plotted in Figure \ref{fig:F1-heat}. 
 The advective heat flux is smaller in magnitude than the conductive flux and asymptotes to zero from negative values as the atmosphere rearranges to an isothermal state.  

\begin{figure} 
\epsscale{1.3}
\plotone{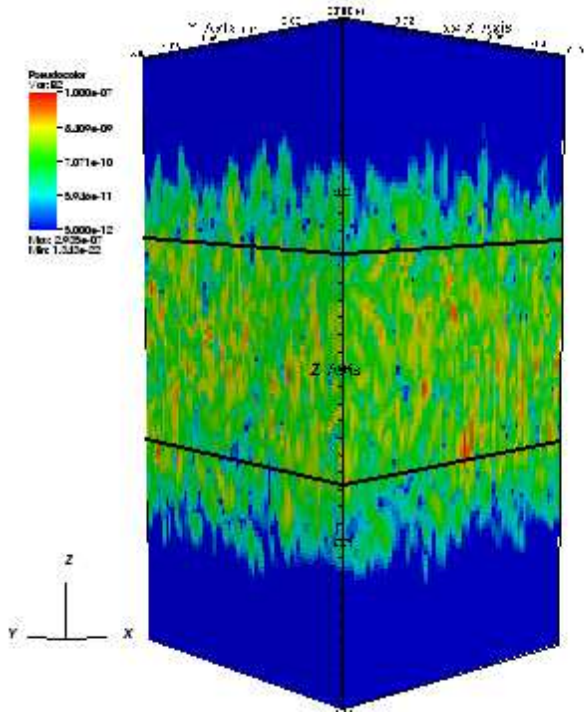}
\caption{Pseudocolor plot of the magnetic energy density at time, $t/\tau_s \approx 3,840$, in run F1.  Note that the unstable region extends from $0.067 \le z \le 0.13$ as denoted by the solid lines.  The initial magnetic energy density is $B^2 = 10^{-10}$ in the unstable region and zero elsewhere.  The highly subsonic turbulence drives little penetrative convection into the stable region.}\label{fig:F1-B2}  
\end{figure}

The saturation of Case A can be described by a quick relaxation of the temperature gradient followed by modest magnetoconvection until the dynamo has increased the magnetic field until it is in equipartition with the turbulent convective motions.  The convective velocities are highly subsonic and the magnetic field remains principally confined to the unstable region.  More quantitatively, the time for a fluid element to cross the unstable layer at the RMS velocity is $L/\langle v\rangle \approx 1.8\times 10^4 \tau_s$, which is far longer than the time it takes for the simulation to saturate.  Figure \ref{fig:F1-B2} shows a visualization of the magnetic energy density at time, $t/\tau_S\approx 3,840$.  The buoyant plumes are largely confined to the stable region with little flux pumping to the stable region.

Runs N1, N4, and N5 also saturate with a case A mechanism.  N2 is the exact same as the fiducial case except it is initialized with a net magnetic flux.  The isotropic thermal conductivity in regions 1 and 3 and the anisotropic conductivity in region 2 are a factor of five higher in runs N4 and N5 than run F1, thus facilitating a comparison of the role of thermal conductivity.  Runs N4 and N5 differ in that the latter has a net magnetic flux threading the domain; whereas, the former has zero net magnetic flux.  It is interesting to note that in two dimensions, the zero net magnetic flux condition requires the magnetic energy density to decay away by Cowling's anti-dynamo theorem \citep{cowl34}.  In three dimensions, magnetic dynamo action is permitted, and runs F1 and N4 show a sustained increase in the magnetic field.  By examining Table \ref{tab:satprop} one can compare the runs at the fiducial time of $t/\tau_s = 2,050$.  At this time, the temperature gradient is essentially fully relaxed, but the magnetic dynamo has not had quite enough time to fully amplify the magnetic energy density to equipartition with the kinetic energy.  

Comparing runs F1 and N4, we see that the higher conductivity only marginally speeds the increase of the magnetic energy density.  A slightly larger increase is seen in the saturation of the temperature gradient (10.7\% versus 15.8\%).  The higher conductivity does result in an almost three-fold increase in the RMS Mach number.  On the whole, the uniform increase in conductivity does not qualitatively change the saturation mechanism.  The saturated heat flux is very similar with $\langle Q\rangle/\widetilde{Q} \approx 44\%$.

Comparing runs F1 versus N1 and N4 versus N5, we see that a net magnetic flux threading the domain produces a marked increase in the strength of the magnetic dynamo action bringing the magnetic energy closer to equipartition faster in the runs with the net flux.  The magnitude of the convective velocities remains the same in both cases.  The degree of saturation of the temperature gradient is the same, but the average angle of the magnetic field with respect to the horizontal is greater in the net flux case.  At this early stage of saturation, the advective heat flux is often negative as the atmosphere reaches the new equilibrium.    

\subsection{Case B: Reorientation of Magnetic Field Geometry} \label{subsec:saturation:F3}
\begin{figure*} 
\epsscale{0.70}
\plotone{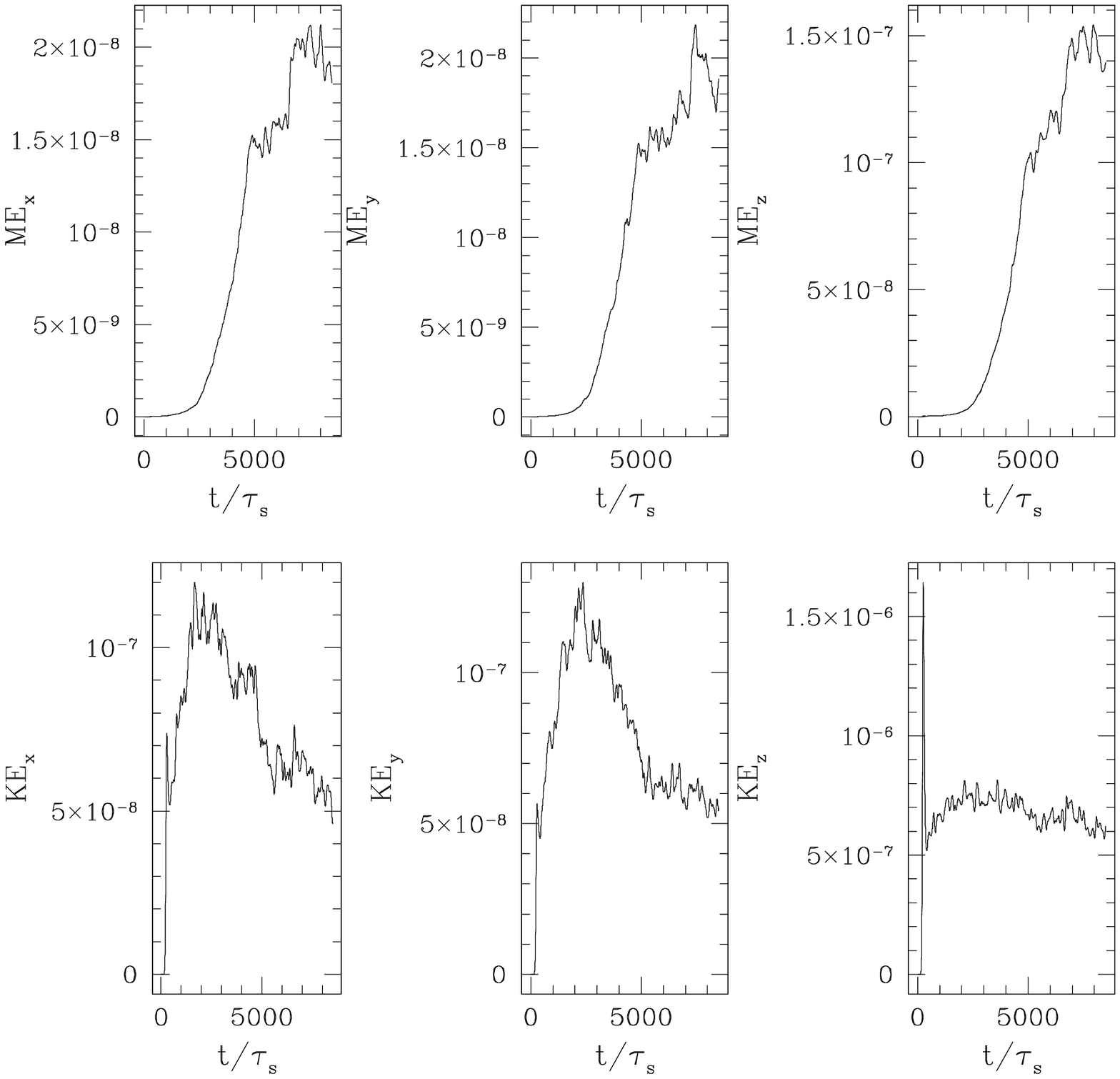}
\caption{Time evolution of the components of the volume-averaged magnetic and kinetic energy density for the entire domain for Run F3.}\label{fig:F3-time}
\end{figure*}

\begin{figure} 
\epsscale{1.2}
\plotone{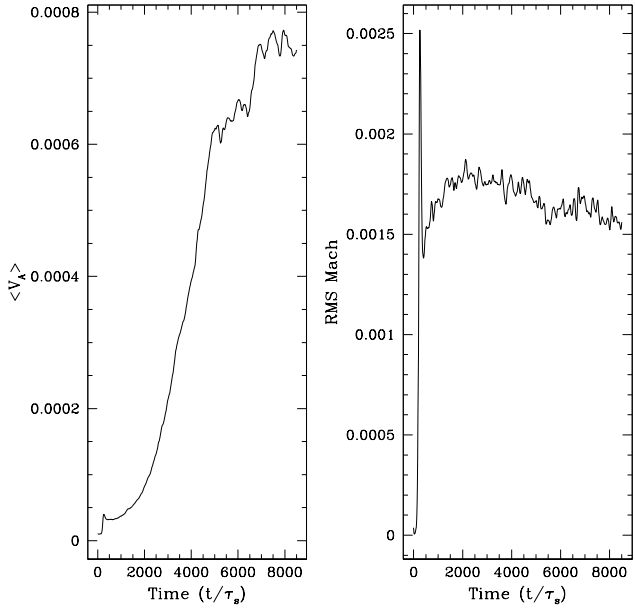}
\caption{Evolution of the Alfv\'{e}n velocity and RMS Mach number of run F3 averaged over the central MTI-unstable region.  The two are almost in equipartition.}\label{fig:F3-midavg}
\end{figure}

We proceed to discuss a very different saturated state for the MTI, exemplified by run F3.  In this run, the conductivity in the isotropic, stable layers is an order of magnitude larger than the anisotropic conductivity.
Figure \ref{fig:F3-time} shows the time evolution of the magnetic energy density and kinetic energy density over the entire computational domain.  There is a very strong amplification of the magnetic energy density in the unstable domain by a factor of approximately $5.3\times 10^{3}$ or $7.8\times 10^{3}$ for $\langle B\rangle^2$ and $\langle B^2 \rangle$, respectively, as defined in Equation \ref{eqn:bamp}.  Figure \ref{fig:F3-midavg} shows the corresponding time evolution of the Alfv\'{e}n velocity and the RMS Mach number.  The Mach number of the turbulence has a brief spike corresponding to the initial growth of the instability before reaching a very stable $1.8\times 10^{-3}$.  The Alfv\'{e}n velocity is within a factor of two of the turbulent velocity with slight variation depending on which average of the magnetic field is utilized.  


\begin{figure} 
\epsscale{1.2}
\plotone{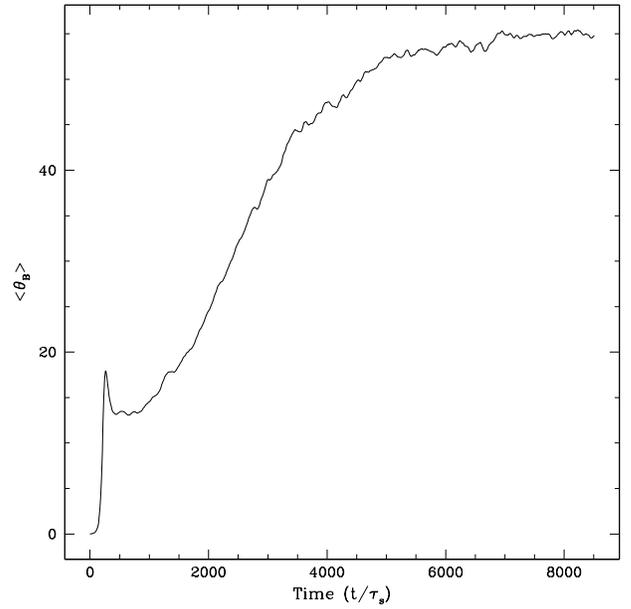}
\caption{Evolution of the average angle of the magnetic field in the central region with respect to the horizontal in degrees for run F3.  The sharp initial peak is followed by a steady increase to the quite large asymptotic value of $\langle\theta_B \rangle\approx 55\degree$ }\label{fig:F3-btheta}  
\end{figure}

\begin{figure} 
\epsscale{1.2}
\plotone{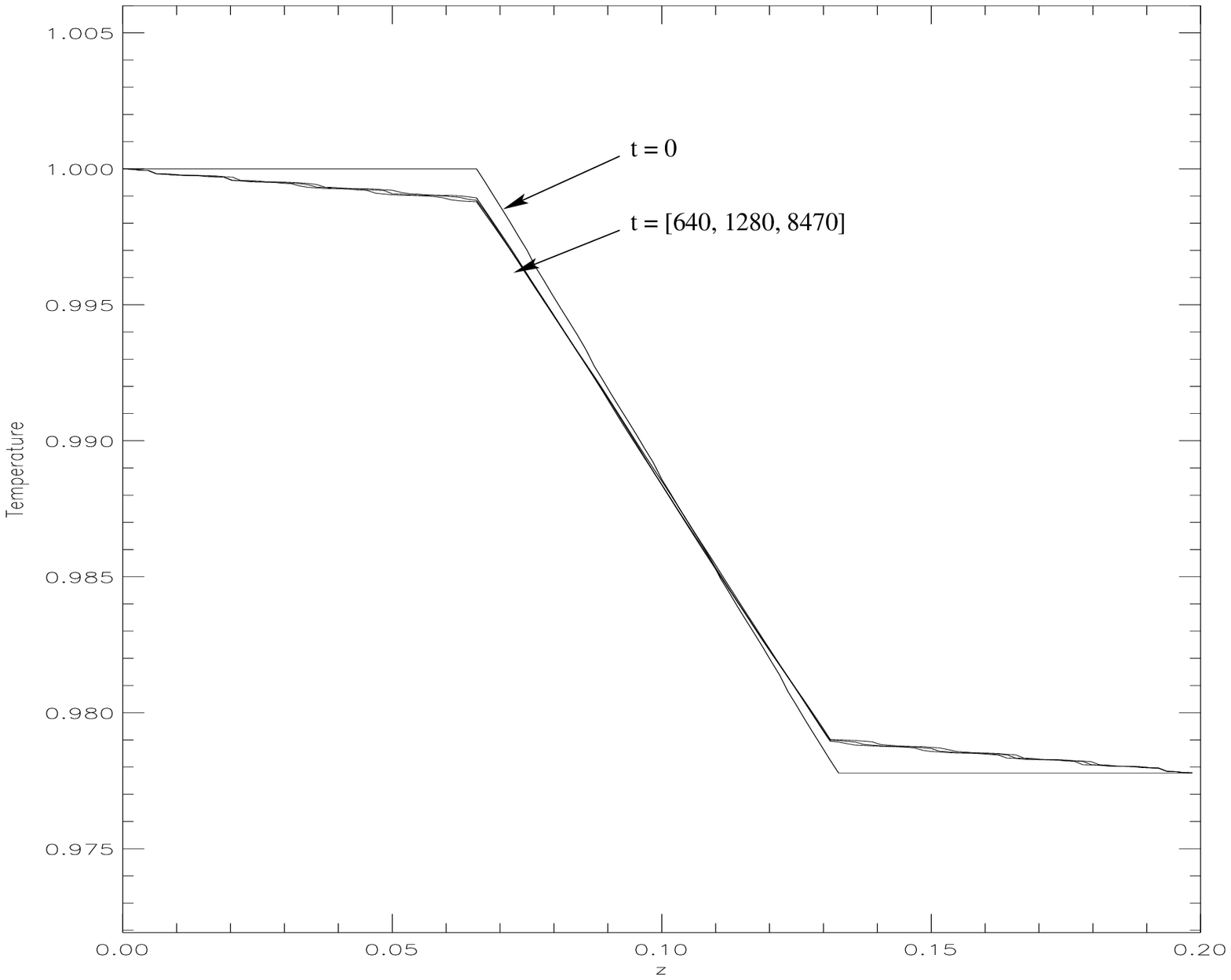}
\caption{Vertical profile of the horizontally-averaged temperature profile in run F3 at various times.  Times are labeled in units of $\tau_s$.  The temperature evolution happens almost entirely in the first 650 sound crossing times or so.  The three times in brackets are all essentially collinear profiles.}\label{fig:F3-tavg}
\end{figure}

Up until this point in the discussion, the magnetoconvection in run F3 simply resembles a more energetic version of run F1.  The differences are exposed by examing the evolution of the angle of the magnetic field with the horizontal, shown in Figure \ref{fig:F3-btheta}, 
and the evolution of the horizontally-averaged, vertical temperature profile, shown in Figure \ref{fig:F3-tavg}.  It is obvious that the temperature profile evolves quickly to its final value, which deviates only slightly from the initial profile, i.e. approximately $91\%$ of the initial temperature gradient is the asymptotic value.  In contrast, the average angle of the magnetic field evolves quite dramatically to a primarily vertical field with $\langle\theta_B \rangle\approx 55\degree$.  In this case, which we call Case B, the MTI saturates not by temperature gradient relaxation, but instead by a reorientation of the magnetic field to a primarily vertical field.

\begin{figure*} 
\epsscale{0.6}
\plotone{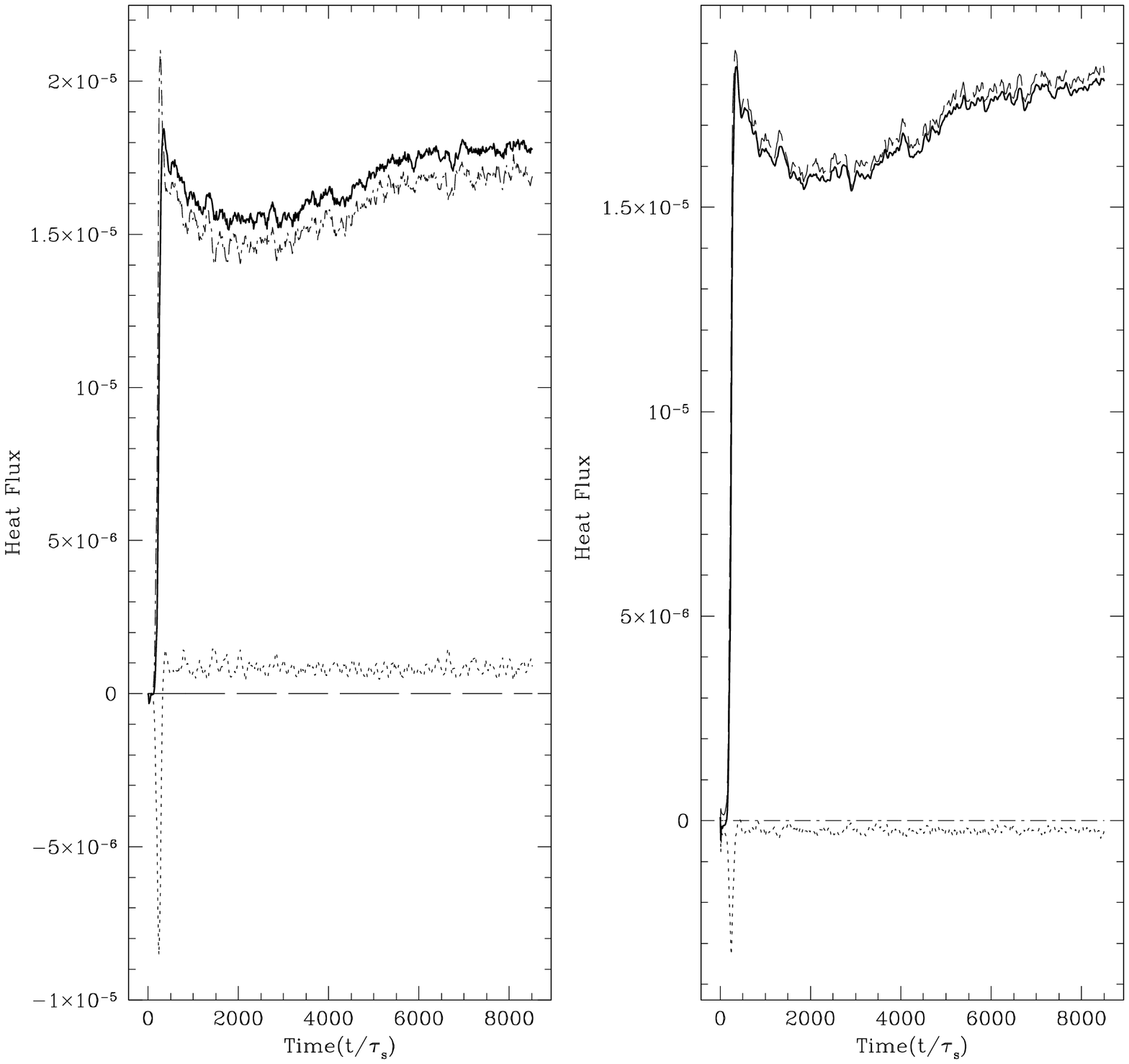}
\caption{Time evolution of the horizontally-averaged heat flux at the midplane (\textit{left panel}) and at 80\% height (\textit{right panel}) of the simulation domain in run F3.  The total heat flux (\textit{solid line}) is subdivided into conductive (\textit{dot-dash line}), radiative (\textit{long dashed line}), and advective (\textit{dotted line}) components.  Both the time-average conductive and advective heat fluxes are positive, but the conductive flux dominates. }\label{fig:F3-heat}  
\end{figure*}

We turn our attention to its transport properties.  Figure \ref{fig:F3-heat} plots the time evolution of
the vertical components of these three heat fluxes and the total heat
flux at the midplane, and at 80\% of the height of the domain in Run F3.  At the midplane, the Coulombic heat flux is the dominant component of the heat flux with a smaller positive contribution from the advective heat flux.  
It is interesting to estimate the efficiency of the MTI in transporting heat.  We can calculate the components of the heat flux \textit{a posteriori} from our data.  Using Equation \ref{eqn:F1-anisoQ} to estimate the heat flux is not appropriate here since $sin\langle\theta_B \rangle \neq \langle sin \theta_B \rangle$ for large angles.  The ideal heat flux for comparison purposes is essentially purely isotropic transport across the unstable layer, which is 
\begin{equation}
\widetilde{Q} = - \chi_{\textrm{aniso},0}\left(\frac{\dif T}{\dif z}\right)_{\textrm{initial}} \approx (1\times10^{-4})(0.333) \approx 3.3 \times 10^{-5}.
\label{eqn:F3-idealflux}
\end{equation}
Thus, the essentially constant Coulombic flux of $1.6\times 10^{-5}$ is almost exactly half of the fiducial heat flux.  This fact leads us to conclude that the MTI can be very efficient at heat transport.  

One fact that bears further elucidation is the small time-average of the advective component of the heat flux.  With the exception of gravity, our domain is essentially symmetric from top to bottom.  In addition, the box size is small compared to the scale height, so our simualtions are intrinsically local in nature.  In a real plasma of interest, the Spitzer conductivity, which varies as $\kappa_{sp} \propto T^{5/2}$, would have a substantial variation in thermal conductivity in the vertical direction.  Such a conductivity variation, combined with a potential variation in optically-thin cooling and a global temperature profile, could produce a larger net advective flux.  Such a realistic configuration will be pursued in future work.

\begin{figure}[b] 
\epsscale{1.3}
\plotone{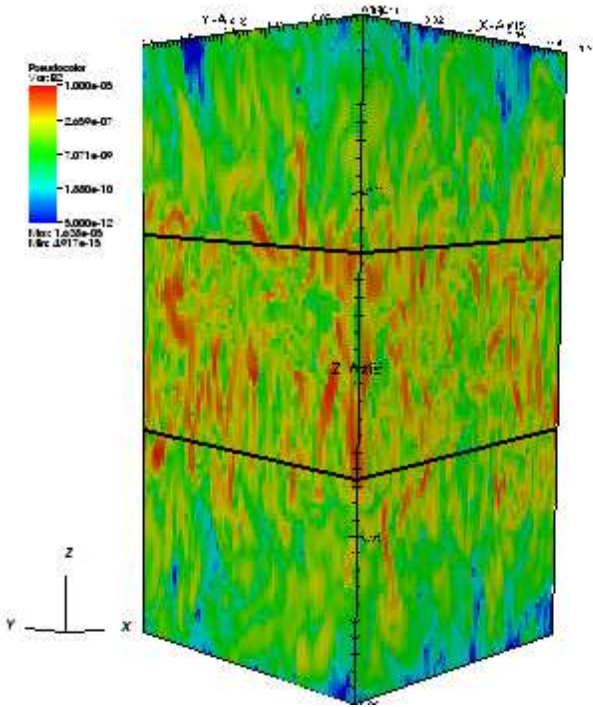}
\caption{Pseudocolor plot of the magnetic energy density at time, $t/\tau_s \approx 4,700$, in run F3.  Note that the unstable region extends from $0.067 \le z \le 0.13$ as denoted by the solid lines.  The initial magnetic energy density is $B^2 = 10^{-10}$ in the unstable region and zero elsewhere.  Note the plumes due to penetrative convection and the storage of magnetic flux in the stable layers.}\label{fig:F3-B2}  
\end{figure}

We now discuss one further aspect of run F3 related to the magnetoconvection.  Figure \ref{fig:F3-B2} shows a visualization of the magnetic field energy density in the simulation at time, $t/\tau_s \approx 4,700$.  At this intermediate time, the maximum magnetic energy density exceeds $B^2 = 10^{-5}$, five orders of magnitude larger than the initial value, for regions colored red in the plot.  In addition, one can see buoyant plumes penetrating into the stable region as a result of convective overshoot.  Much magnetic field has now been stored in the stable layers which initially had zero magnetic field (colored blue in the plot).  The phenomena of penetrative convection and overshooting has been thoroughly studied in the magnetoconvection literature of stellar atmospheres \citep{tob01, bru02}.  In comparison to this run, Case A saturation has far less penetrative convection due to the weaker magnetoconvection (compare to Figure \ref{fig:F1-B2}).  

Finally, we compare run F3 to the other Case B runs (N2, N3, and N6).  N2 has the same parameters as F3, but has a net magnetic flux.  N6 has a fivefold increase in both components of the thermal conductivity and a net flux.  We can compare the fiducial time of $t/\tau_s = 2,050$, which is not yet fully saturated for either simulation.  As we would expect based on our prior discussion, runs N2 and N6 have amplified the magnetic field modestly more in the same amount of time; however, the temperature gradient saturation is similarly minimal in both cases.  Run N3 is a check for convergence having doubled the resolution in each direction.  As would be expected for a higher resolution simulation, run N3 is approaching saturation more quickly than the lower resolution fiducial case.  The asymptotic behavior though results in very similar magnetic field amplification, levels of turbulence, temperature profile saturation, and most importantly heat flux.  Thus, we have a check that our fiducial resolution is sufficient.

\subsection{Comparison of the Cases} \label{subsec:saturation:comparison}
Having outlined the two different saturated states of the MTI, we now contrast the two.  Table \ref{tab:satprop} lists the saturated values of Case A (Run F1) and Case B (Run F3).  Case A, temperature gradient relaxation, is quite similar in nature to the 2D simulation with adiabatic boundary conditions in PS.  A small amount of heat flux is supplied to the unstable layer, leading to the isothermal temperature profile. In addition, the field remains primarily horizontal with heat flux dominated instantaneously and in the time average by thermal conduction.  The convection is highly subsonic and little magnetic flux pumping is observed.  The net heat flux is roughly one-half of the fiducial value in the saturated state.  

Case B, on the other hand, has a large heat flux supplied to the unstable layer, which causes the MTI to saturate by re-orienting the field lines in a largely vertical direction.  The level of convection, while still subsonic, is far more vigorous than Case A.  The advective heat flux is consistently positive; however, the total heat flux is still dominated by the Coulombic component.  The convective plumes are able to traverse the unstable layer in approximately 700 sound crossing times, thus leading to a significant amount of convective overshoot and magnetic energy storage in the stable layers.  The net heat flux is roughly one--half of the fiducial value.

In both cases, the magnetic energy density increases through dynamo action until it is roughly in equipartition with the kinetic energy density of the subsonic convection.  It should be noted that the plasma energy density is still dominated by the thermal component.  In both cases, the effective thermal conductivity, as measured by the ratio of the actual heat flux to the fiducial heat flux, is quite high.  

\subsection{The Intermediate Case} \label{subsec:saturation:F2}
\begin{figure*} 
\epsscale{0.60}
\plotone{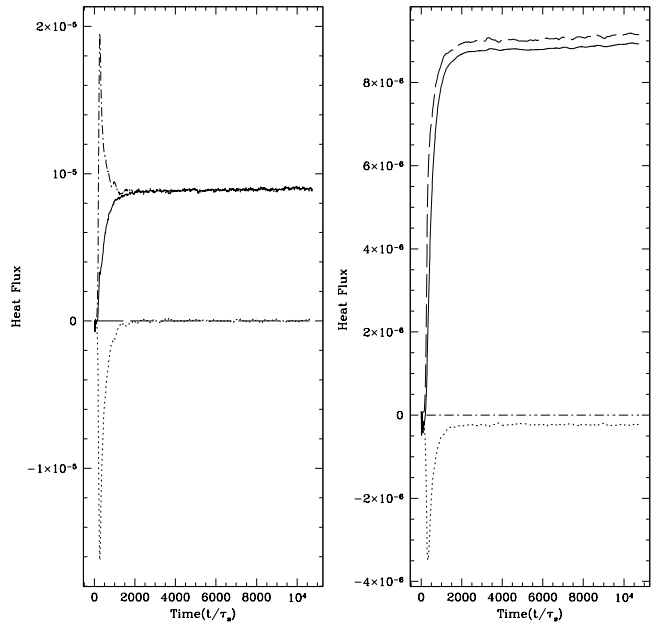}
\caption{Time evolution of the horizontally-averaged heat flux at the midplane (\textit{left panel}) and at 80\% height (\textit{right panel}) of the simulation domain in run F2.  The total heat flux (\textit{solid line}) is subdivided into conductive (\textit{dot-dash line}), radiative (\textit{long dashed line}), and advective (\textit{dotted line}) components.  The conductive flux clearly dominates the advective flux.}\label{fig:F2-heat}  
\end{figure*}

Intermediate to the two cases previously discussed lies run F2 where the thermal conductivity in the middle and outer layers is constant, $\chi_{\textrm{aniso},0} = \chi_{\textrm{iso},0} = 10^{-4}$. 
\begin{figure*} 
\epsscale{0.7}
\plotone{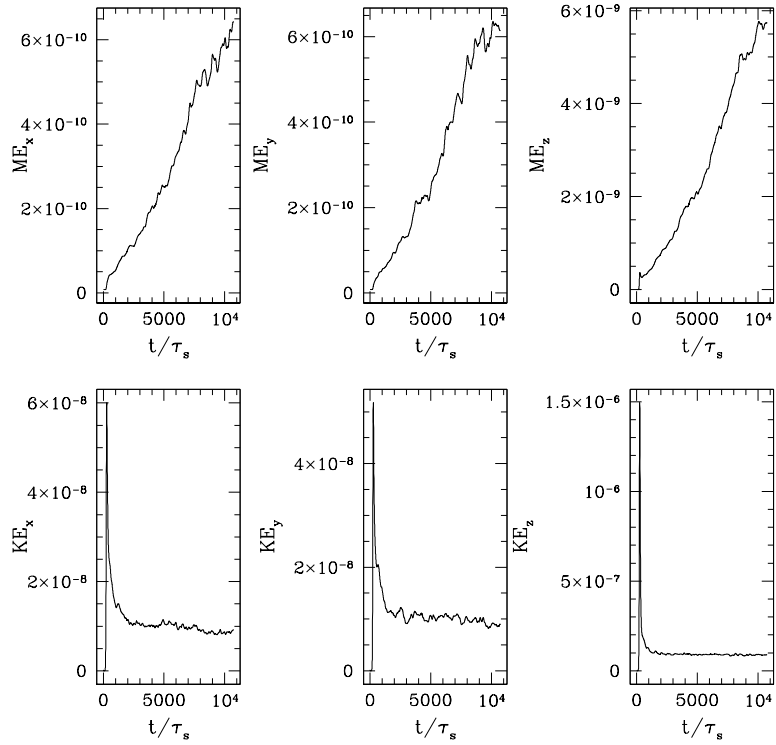}
\caption{Time evolution of the components of the volume-averaged magnetic and kinetic energy density for the entire domain for Run F2.}\label{fig:F2-time}
\end{figure*}
Figure \ref{fig:F2-time} shows the time evolution of the volume--averaged components of the magnetic energy density and kinetic energy density over the entire domain, including the stable regions.  Appreciable amplification of the magnetic energy density is seen, especially in the $z$-component.  The kinetic energy, after an initial transient, settles down to stable constant turbulence with an RMS Mach number of $6.1\times 10^{-4}$ averaged over region 2.  The (essentially) saturated Alfv\'{e}n velocity is $1.4\times 10^{-4}$, thus bringing the magnetic energy density into quasi-equipartition with the kinetic energy density.  The angle of the magnetic field with respect to the horizontal saturates at an angle of approximately $37\degree$.  The heat fluxes for this run are shown in Figure \ref{fig:F2-heat}.  The saturated Coulombic heat flux in both layers is found to be approximately $9 \times 10^{-6}$, slightly less than one-third of the maximum possible value---still consistent with a relatively efficient transport of heat.  The advective heat fluxes, while initially zero, are asymptotically oscillating around zero.  Since we define the fiducial heat flux with respect to the minimum thermal conductivity in the problem (imagine a series of thermal resistors), the intermediate case appears to be less efficient than the other two cases.  In run F1, the minimum conductivity is in the isotropic stable layer, thus causing $\widetilde{Q}$ to be an order of magnitude smaller for it. If we were to instead calculate the fiducial heat flux with only the thermal conductivity of the anisotropic layer, the intermediate case would indeed have a thermal efficiency intermediate to runs F1 and F3.  

\begin{figure} 
\epsscale{1.1}
\plotone{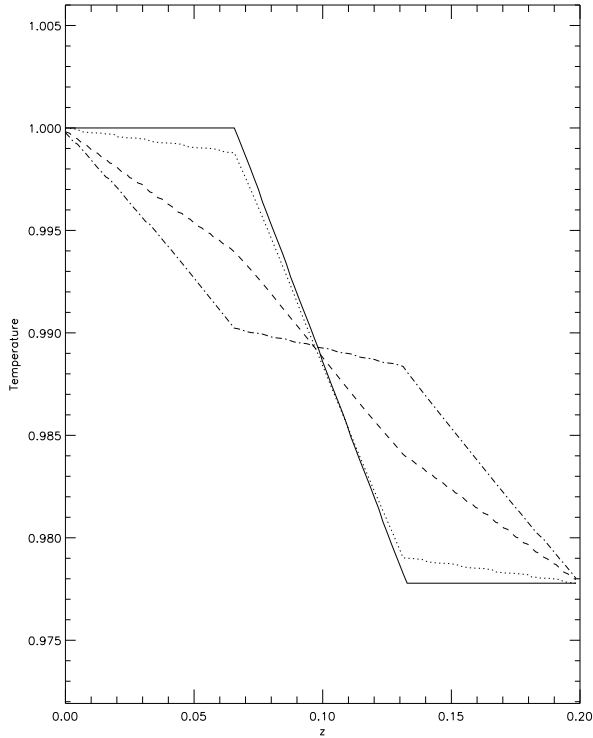}
\caption{Comparison of the horizontally-averaged vertical temperature profiles.  The initial condition is plotted with a solid line.  The saturated temperature profiles of run F3 (\textit{dotted line}), run F2 (\textit{dashed line}), and run F1 (\textit{dash-dot line}) show increasing degrees of relaxation of the temperature profile towards isothermality in the unstable layer.}\label{fig:Tcomp}
\end{figure}

Having outlined three different cases of saturation, we now show that there is a coherent and logical ordering of saturation parameters among them.  As we progress from Case A to Case B, the free parameter we increase is the ratio of isotropic conduction in the outer layers to the anisotropic conduction in the unstable region, $\chi_{\textrm{iso},0}/\chi_{\textrm{aniso},0}$.  This variation is tantamount to increasing the applied heat flux to the MTI-unstable layer.  The resultant saturation properties are evident by simply scanning Table \ref{tab:satprop}.  Firstly, the vigor of the resultant magnetoconvective dynamo, as characterized by amplification of the magnetic field ($\delta\langle B^2\rangle$ and $\delta\langle B\rangle^2$), the RMS Mach number, and the Alfv\'{e}n velocity, increases significantly as we progress from Case A to Case B.  Secondly, the resultant atmospheric state changes smoothly in this progression as evidenced by the increase in the average magnetic field angle.  The degree of relaxation of the temperature gradient, as shown in Figure \ref{fig:Tcomp}, decreases substantially from Case A to Case B.  Yet, the net heat flux, which is a convolution of these two properties, monotonically increases from Case A to Case B. All three cases exhibit quite efficient transport of heat flux compared to the fiducial value.  

\section{Applications} \label{sec:apps}
\subsection{Plasmas in Clusters of Galaxies} \label{subsec:apps:ICM}
The most promising potential application of the MTI is to the intracluster medium of clusters of galaxies. The plasma has a typical density of  $10^{-3}$--$10^{-2} \mathrm{cm}^{-3}$ and a temperature of roughly $1-15$ keV, leading to copious X--ray emission measured by \textit{Chandra} and \textit{XMM-Newton}.  The total cluster luminosities are found to be typically $L_X \sim 10^{43}$ erg/s to $10^{46}$ erg/s in the X--ray.  Most of these clusters appear to be in hydrodynamic equilibrium, as the time since the last major merger is approximately 5 Gyr \citep{pf06}, although a few exceptional examples have relatively disturbed shapes due to recent major mergers, such as Perseus \citep{fab06}, Abell 3667 \citep{vik01}, and Abell 2142 \citep{mark00}.  Estimates for magnetic field strength in clusters range from 1 -- 10 $\mu$G at the center and 0.1 -- 0.7 $\mu$G at a radius of 1 Mpc \citep{bru01, fer99}, values which are typically not dynamically important.   

To understand the potential for the MTI, begin by considering the Spitzer conductivity for a cluster plasma, 
\begin{equation}
\kappa_{\parallel} = 9.2\times 10^{30} n_e k_B\left(\frac{k_B T}{5 \textrm{keV}}\right)^{5/2}\left(\frac{10^{-2}\textrm{cm}^{-3}}{n_e}\right)
\left(\frac{37}{\ln \Lambda_C}\right)\frac{\textrm{cm}^2}{\textrm{s}},
\label{eqn:spitzer}
\end{equation}
where $n_{e}$ is the electron density and $\ln \Lambda_{C}$ is the Coulomb logarithm \citep{chan05}.  Since $\rho_e \ll \lambda_{\textrm{mfp}}$, the perpendicular electron conductivity may be safely ignored, \textrm{e.g.} for a 1 $\mu$G field, $\kappa_{\perp}/\kappa_{\parallel} \sim 10^{-24}$.  The gravitational potential of the cluster is well-approximated by an NFW profile for the dark matter \citep{nfw97}.  Now we can consider the instability criterion for the MTI, Equation \ref{eqn:MTI_instability_criterion1}, for a typical cluster with mass $10^{14}$ $\mathrm{M}_{\odot}$, an NFW scale radius of 500 kpc, and a typical density of $10^{-2}\, \mathrm{g/cm}^3$.  The first term of the instability criterion is a measure of magnetic tension at a given scale and field strength.  For our model cluster, the plasma beta is $\beta \approx 2,000$, thus, the magnetic field is dynamically weak. The second term is set by the product of the cluster gravitational cluster potential ($\rho^{-1}(\dif p/\dif z)=\nabla \Phi$) and the logarithmic derivative of the temperature.  We calculate a reasonable temperature profile by using a fit to cluster structure formation calculations given by \citet{lok02}.  Finally, we assume that outside of the cluster core, conduction is entirely dominated by parallel conduction, thus $\chi'_{C}/\chi'=1$.  Balancing these two terms gives the smallest wavelength, $\lambda_{\mathrm{crit}}$, at which the ICM is marginally unstable to the MTI.  This lengthscale  can be parameterized as
\begin{equation}
\lambda_{\textrm{crit}} = 4.6 \quad \mathrm{kpc} \left(\frac{T}{5 \quad \mathrm{keV}}\right)^{1/2}\left(\frac{2000}{\beta}\right)^{1/2}.
\label{eqn:cluster:lambdacrit}
\end{equation}
Thus all scales larger than 4.6 kpc are unstable to the MTI.  An alternative way of looking at this problem is to estimate what field strength would be required for the largest scale, i.e. the cluster virial radius, to be stable to the MTI.  For our model cluster this requires $\beta \approx 0.16$, which corresponds to magnetic field strengths in excess of 110 $\mu$G.  Typical measurements of cluster field strengths by the various mechanism of synchrotron emission, Faraday rotation, or inverse Compton emission never approach this large value \citep{ct02}.  

In addition, we can comment on the temperature profiles of clusters.  The previously mentioned structure formation calculations assume a $\Lambda$CDM cosmology and predict a moderately steep, monotonically decreasing temperature profile of the intracluster plasma \citep{lok02}.  Recent observations with \textit{XMM-Newton} have inferred a temperature profile that is flatter than the predictions from structure formation \citep{pratt06}. It may be that the observed temperature profiles are indicative of the saturated state of the MTI.  The MTI also provides for a dynamo mechanism to amplify a small seed magnetic field through magnetoconvection.  

A number of authors have considered the effect of thermal conduction on the temperature profile of clusters of galaxies using an effective conductivity across tangled field lines in a turbulent medium \citep{rr78, cc98}.  More recent efforts have revised this calculation to include a spectrum of correlation lengths, resulting in an effective conductivity that is much closer to the value of the full Spitzer conductivity \citep{nm01, zn03}.  In this case the turbulence which tangles field lines is externally driven.  However, the ICM is unstable to the MTI, and therefore the turbulence, the tangling of the field lines, and the resulting heat flux are determined by the same underlying driving mechanism.  Computing the temperature profile and saturated heat flux self-consistently in a typical ICM is an important goal of future efforts.  

\citet{chan05} has constructed an analytical model that includes the effects of anisotropic cosmic ray pressure in addition to the thermal conduction in the MTI model.  It is worth noting that non-thermal particles may also contribute to the structure and dynamics of the ICM. In addition, \citet{cd06} have proposed that a cosmic ray version of the MTI could be at play in the core of the cluster, where a radial gradient of cosmic ray pressure (from an AGN for example) can play the role the temperature gradient does in the MTI.  This heat transport in the core is a potential mechanism to obviate the need for cooling flows. 

\subsection{Other Applications} \label{subsec:apps:other}
Two other potential applications of the magnetothermal instability have also been identified.  First, neutron star atmospheres have large temperature and pressure gradients due to a surface gravity of $g \approx 2\times 10^{14}\, \textrm{cm/s}^2$, giving a scale height of around 10 cm.  The temperature gradient must be large to drive the MTI in the presence of large magnetic fields.  For example, millisecond radio pulsars typically have magnetic fields in the $10^8$ -- $10^9$ G range \citep{bhat95}.  In a region near the equator and inside of the photosphere, there is the potential for the MTI to be unstable.  Here the primary limiting factor is the term $\chi'_{C}/\chi'$ from the instability criterion.  The atmosphere is a semi-degenerate, semi-relativistic plasma.  Deep in the atmosphere, the Larmor radius becomes comparable to the mean free path.  Near the surface of the atmosphere, radiative transport primarily due to Brehmsstrahlung becomes important.  Both of these properties suppress the MTI due to isotropic transport.  Between these two regions, though, the neutron star atmosphere may be MTI-unstable.  Due to the complex nature of the physics, this supposition will require detailed modeling of the neutron star atmosphere to assess with certainty.  

A second potential application is to Radiatively Inefficient Accretion Flows (RIAFs).  Observations of the electron mean free path in Sgr A$^*$ in our Galactic Center show that $\lambda_{\textrm{mfp}} \sim 10^5$ -- $10^6$ Schwarschild radii.  Thus, the plasma is strongly collisionless.  Recent analytical work by \citet{jq06} shows that including electron thermal conduction can suppress the accretion rate by 1 -- 3 orders of magnitude from the Bondi rate---a first step in explaining the very low luminosity of Sgr A$^*$.  This analysis is static, however, and cannot fully account for the resultant buoyant motion driven by the thermal conduction.  \citet{shq03} have developed Landau fluid closures that describe kinetic MHD appropriate to this regime.  Fully collisionless shearing box simulations performed by \citet{shar06} actually lead to a modestly increased $\alpha$-parameter compared to typical MRI turbulence in an MHD simulation.  This increase in $\alpha$ is due to anisotropic pressure effects in the fully collisionless treatment.  Such an increase is consistent with the magnetoviscous instability in the shearing box \citep{bal04}; however, it cannot account for  observations of much lower accretion rates for Sgr. A$^*$.  As a result, global simulations of collisionless accretion are called for that include the temperature stratification of the flow.

\section{Summary and Future Work} \label{sec:conclusion}
Many astrophysical plasmas fall into a low density, weakly magnetized regime, in which transport of heat and momentum due to thermal conduction and viscosity takes place primarily along magnetic field lines.  One would expect an atmosphere with $\dif S/\dif z >0$ to be completely stable to convection.  Yet, the magnetothermal instability destabilizes this atmosphere if $\dif T/\dif z < 0$.  We have performed fully three-dimensional time-dependent MHD simulations of the nonlinear regime of this instability.  

The nonlinear evolution of the MTI in 3D with different parameters shows several common features.  Exponential growth in the linear growth rate produces a spike in all the components of the kinetic and magnetic energy densities, and a rapid rearrangement of the temperature profile to the final saturated temperature gradient. After this initial transient, the atmosphere enters a magnetoconvective phase in which the fluid motions reach a steady level of subsonic turbulence.  Simultaneously, the magnetic field is amplified in a dynamo until the magnetic energy density is in approximate equipartition with the kinetic energy density.  

As was outlined in \S \ref{sec:saturation}, we have identified two different saturated states for the MTI in 3D.  These mechanisms differ proximately in the strength of the isotropic thermal conduction in the stable layers relative to the strength of the anisotropic thermal conduction in the MTI-unstable region.  This adjustable parameter allows an exploration of the parameter space for a range of driven heat flux boundary conditions.  In Case A: temperature gradient relaxation, the isotropic heat flux is small and the unstable layer saturates in a way reminiscent of adiabatic simulations from 2D.  In the saturated state, the temperature profile is almost isothermal and the magnetic field is still primarily horizontal in orientation.  In Case B, the isotropic heat flux is large and the unstable layer saturates in a significantly different way.  The temperature gradient relaxes only slightly; however, the magnetic field rearranges itself into a primarily vertical configuration to carry the applied heat load.  This variation of the isotropic heat flux changes the vigor of the magnetoconvection, resulting in higher RMS Mach numbers and higher magnetic field amplification as we approach Case B.  In all cases, though, we find the instability reaches a natural quasi-equipartition between subsonic turbulent motions and magnetic energy density.

A key conclusion from this work is that the magnetic field can no longer be thought of a passive contaminant.  We began with an atmosphere that was convectively stable to the Schwarschild criterion with a weak magnetic field.  We can quantify ``weak'' in terms of the plasma beta, $\beta = 4\pi p/ B^2$, the ratio of thermal to magnetic pressure.  For our initial choice of magnetic field in the simulations, $\beta = 10^{10}$.  Thus, conventional wisdom leads to considering the magnetic field as dynamically irrelevant---a passive contaminant simply frozen into the background fluid and transported by purely hydrodynamic motions.  Yet, the illustrative example of Case B shows that this reasoning is patently incorrect in many astrophysical plasmas.  The transport of heat along the magnetic field triggers the magnetothermal instability which is capable of reordering and amplifying the magnetic field, generating magnetoconvection, and transporting heat.  The dynamically weak magnetic field now completely dominates the resulting dynamics.

The most immediate application for future study is to the intracluster medium of clusters of galaxies.  A fully three-dimensional MHD simualtion including realistic Spitzer thermal conductivity and radiative cooling is required.  In addition, the effects of conduction on radiatively inefficient accretion flows and convection dominated accretion flows must be investigated in the context of MRI turbulence in global MHD simulations.  

\acknowledgements
We thank Tom Gardiner for his contributions to the Athena code used for this study and Steve Balbus for useful discussions.  IP acknowledges support from the Department of Energy Computational Science Graduate Fellowship.  JS acknowledges support from DoE grant DE-FG52-06NA26217.  Simulations were performed on computational facilities supported by NSF grant AST-0216105 and an IBM BlueGene/L Supercomputer at Princeton University.

\end{document}